# High current density electrical breakdown of TiS₃ nanoribbon-based field-effect transistors

*Aday J. Molina-Mendoza,* [†]* *Joshua O. Island,* [†] *Wendel S. Paz, Jose Manuel Clamagirand, Jose Ramón Ares, Eduardo Flores, Fabrice Leardini, Carlos Sánchez, Nicolás Agraït, Gabino Rubio-Bollinger, Herre S. J. van der Zant, Isabel J. Ferrer, J.J. Palacios, and Andres Castellanos-Gomez.* *

[†]*These authors contributed equally*

Dr. Aday J. Molina-Mendoza[1], Wendel S. Paz[1], Prof. Nicolás Agraït[1,2,3], Prof. Gabino Rubio-Bollinger[1,3] and Prof. J.J. Palacios[1]

[1]Departamento de Física de la Materia Condensada, Universidad Autónoma de Madrid, Campus de Cantoblanco, E-28049, Madrid, Spain.

[2]Instituto Madrileño de Estudios Avanzados en Nanociencia (IMDEA-nanociencia), Campus de Cantoblanco, E-18049 Madrid, Spain.

[3]Condensed Matter Physics Center (IFIMAC), Universidad Autónoma de Madrid, E-28049 Madrid, Spain

E-mail: aday.molina@uam.es

Dr. Joshua O. Island[4[+]] and Prof. Herre S. J. van der Zant[4]

[4]Kavli Institute of Nanoscience, Delft University of Technology, Lorentzweg 1, 2628 CJ Delft, The Netherlands

[+]Current Address: Department of Physics, University of California, Santa Barbara, California 93106-6105 USA

Dr. Jose Manuel Clamagirand[5], Prof. Jose Ramón Ares[5], Eduardo Flores[5], Prof. Fabrice Leardini[5], Prof. Isabel J. Ferrer[5,6] and Prof. Carlos Sánchez[5,6]

[5]Materials of Interest in Renewable Energies Group (MIRE Group), Dpto. de Física de Materiales, Universidad Autónoma de Madrid, Campus de Cantoblanco, E-28049 Madrid, Spain

[6]Instituto de Ciencia de Materiales "Nicolás Cabrera", Campus de Cantoblanco, E-28049 Madrid, Spain.

Dr. Andres Castellanos-Gomez[2]

[2]Instituto Madrileño de Estudios Avanzados en Nanociencia (IMDEA-nanociencia), Campus de Cantoblanco, E-28049 Madrid, Spain.

E-mail: andres.castellanos@imdea.org

## Abstract

The high field transport characteristics of nanostructured transistors based on layered materials are not only important from a device physics perspective but also for possible applications in next generation electronics. With the growing promise of layered materials as replacements to conventional silicon technology, we study here the high current density properties of the layered material titanium trisulfide ($TiS_3$). We observe high breakdown current densities up to $1.7 \cdot 10^6$ A cm$^{-2}$ in $TiS_3$ nanoribbon-based field-effect transistors which are among the highest found in semiconducting nanomaterials. Investigating the mechanisms responsible for current breakdown, we perform a thermogravimetric analysis of bulk $TiS_3$ and compare the results with density functional theory (DFT) and Kinetic Monte Carlo calculations. We conclude that oxidation of $TiS_3$ and subsequent desorption of sulfur atoms plays an important role in the electrical breakdown of the material in ambient conditions. Our results show that $TiS_3$ is an attractive material for high power applications and lend insight to the thermal and defect activated mechanisms responsible for electrical breakdown in nanostructured devices.

## Introduction

Two-dimensional (2D) materials have been proposed as prospective candidates to replace conventional semiconductors in electronic devices owing to their attractive properties such as high on/off ratios when integrated in field-effect devices, high carrier mobilities and minimal influence from short-channel effects.[1, 2] Nevertheless, one of the central considerations in the continued miniaturization of solid state devices is the increase in current density passing through the device that leads to electrical breakdown due to Joule heating or related phenomena.[1, 3, 4] Therefore, nanomaterials that can withstand high current densities are highly sought after for electronic applications and the study of their thermal stability is crucial to improved performance and general device stability.

Apart from the transition metal dichalcogenides (TMDCs), such as $MoS_2$, which have already proven to be efficient materials for electronic applications,[5] other materials have arisen as promising candidates for replacing the common semiconductors in electronics and optoelectronics.[6] Titanium trisulfide ($TiS_3$) is a material from the transition metal trichalcogenides (TMTC) family which possess a relatively narrow bandgap of 1 eV,[7] and can be isolated into mono- or few-layer nanoribbons and nanoplates.[8, 9] $TiS_3$ nanoribbons have already shown remarkable field-effect characteristics,[8, 10] ultrahigh photoresponse to NIR light, predicted carrier mobilities up to $10^4$ cm² V⁻¹s⁻¹ and in-plane anisotropy.[11]

Here we present a study of the electrical breakdown and thermal stability of $TiS_3$. We find that the maximum current densities in $TiS_3$ field effect transistors (FETs) can reach $1.7 \cdot 10^6$ A cm⁻² which rivals state-of-the-art FETs based on other nanostructured semiconductors. Through the 1D heat equation we estimate the temperature at which the breakdown occurs (400 ± 50 °C) and find that the electrical breakdown is not strictly due to Joule heating but is most likely triggered by crystal defects (vacancies). Thermogravimetric analysis of bulk $TiS_3$ suggests an oxygen activated desorption of sulfur which leads to vacancies that could be responsible for triggering the electrical breakdown in $TiS_3$ FETs. Through DFT calculations we analyze the energies required for vacancy formation in $TiS_3$ crystals and find that an oxygen activated sulfur desorption is the most plausible scenario. Altogether, the estimated temperature at electrical breakdown in $TiS_3$ FETs is consistent with an oxygen-activated rapid desorption of sulfur observed in bulk material and corroborated through DFT calculations of vacancy formation.

## Results and discussion

$TiS_3$ is a layered material belonging to the monoclinic $ZrSe_3$-type crystal structure shown in **Figure 1a** and 1b.[12] It possesses a chain structure of trigonal prisms stacked one of top of the other in which each Ti atom is coordinated to six S atoms at the corners and another two S atoms in the neighbor chains. The prisms are connected, forming a chain along the *b* axis, while they are bonded to neighboring chains, forming a sheet-like unit. The *b* axis chains make the material highly anisotropic.[13] The layers are stacked on top of each other along the *c* axis by van der Waals bonds, allowing to mechanically exfoliate the bulk material to a few-layer structure.

The synthesis of bulk $TiS_3$ is accomplished by a solid-gas reaction between Ti powder (~1 g, Goodfellow, 99.5% purity) and S gas produced by heating S powder (~1.4 g, Merck, 99.99% purity) inside vacuum sealed ampoules at 500 °C with an S/Ti molar ratio = 3 during 10 days. Figure 1c shows photographs of the resulting $TiS_3$ ribbons within the ampoule after growth. In Figure 1d we show a scanning electron microscopy (SEM) image of the $TiS_3$ powder where ribbons with different thicknesses and lengths can be clearly identified. The quality and crystal structure of the synthetized material is confirmed by Raman spectroscopy (Figure 1e), where vibrational modes ($A^1_g$) corresponding to $TiS_3$ can be distinguished (175 cm⁻¹, 300 cm⁻¹, 370 cm⁻¹, 557 cm⁻¹).[9, 14]

We fabricate field-effect transistors (FET) based on few-layer $TiS_3$ nanoribbons obtained by mechanical exfoliation of bulk material (with thicknesses ranging between 15 nm and 35 nm, ~ 17 to 40 layers). The exfoliated nanoribbons are then transferred to a SiO$_2$ substrate (285 nm thickness) thermally grown on a highly *p*-doped Si substrate by a deterministic transfer method.[15] Briefly, a polydimethylsiloxane (PDMS) stamp (Gelfilm from GelPak®) is placed on bulk TiS$_3$ and peeled off fast, removing several nanoribbons with different thicknesses which are cleaved again by peeling with another PDMS stamp. The stamp is then placed on the SiO$_2$/Si(p$^+$) substrate and removed slowly, leaving the exfoliated material on the surface. We use standard e-beam lithography and lift-off procedures to define the metallic contacts (5 nm Ti/50 nm Au). A scanning electron microscopy (SEM) image of representative devices is shown in **Figure 2a** and b. The electrical characterization of the devices was carried out in a high vacuum (pressure lower than 10$^{-5}$ mbar) probe station. The current-voltage characteristics ($I_{ds}$-$V_{ds}$) of one device is shown in Figure 2c and the transfer curve (current passing through the device while sweeping the back gate voltage for a fixed $V_{sd}$, $I_{sd}$-$V_g$) shows n-type behavior with ~10$^3$ on/off ratio (Figure 2d) and off-state current of ~530 pA. The lower on/off ratio is most likely a result of midgap electronic states from sulfur vacancies.[8] From the transfer curve we extract the FET mobility using the following equation:

$$\mu = \frac{L}{W C_i V_{ds}} \frac{\partial I_{ds}}{\partial V_g} \qquad (1)$$

Where $L$ and $W$ are the channel length and width, respectively, and $C_i$ is the capacitance per unit area to the gate electrode. Using equation (1) and a parallel plate capacitor model ($L$ = 500 nm, $W$ = 100 nm, $C_i = \varepsilon_{SiO2}/d_{SiO2} = 1.23 \cdot 10^{-4}$ F m$^{-2}$, with $d_{SiO2}$ the thickness of the SiO$_2$ substrate), we estimate a two-terminal mobility of ~ 1 cm$^2$ V$^{-1}$ s$^{-1}$, similar to the one reported for TiS$_3$ nanoribbons FETs.[8] Here we have to take into account that the measurement is done with two terminals, what is known to give an underestimated value for the mobility. In addition, we estimate a contact resistance for one device of 8.40 MΩ/μm using the transfer length method (TLM), see Supporting Information for details. This value is comparable to MoS$_2$ transistors with gold contacts.[16]

The electrical breakdown of TiS$_3$ nanoribbons was studied in air at room temperature in order to investigate the FETs in standard operating conditions for electronic devices, nevertheless, one device was also measured in vacuum, obtaining similar results (we address the reader to the Supporting Information for details). The procedure used for determining the electrical breakdown of the devices goes as follows: the back-gate voltage is set to +40 V to turn the device to the ON state, i.e., to reduce the channel resistance. A voltage is applied at one of the two electrodes (source), while the current is measured at the other electrode (drain), establishing a drain-source voltage which is slowly swept (208 mV s$^{-1}$ in steps of 6 mV) from 0 V up to the breakdown voltage, usually between 14 and 20 V (**Figure 3a**). The breakdown current and voltage values can be easily identified from the sharp drop of the current to 0 A (green circles in Figure 3a). Atomic force microscopy (AFM) topographic images of one representative device before and after electrical breakdown are shown in Figure 3b and 3c, respectively, where the difference in thickness before (18 nm) and after (4 nm) breakdown is a sign of the degradation of the material. The highest current density at breakdown measured in our TiS$_3$ FETs is 1.7 · 10$^6$ A cm$^{-2}$, which is among the highest reported when directly compared with the breakdown current density measured for other nanomaterials (multilayer MoS$_2$,[17] multilayer graphene,[18] monolayer graphene,[19] Cu,[20] TiO$_2$,[21] SnO$_2$,[21] Si,[22] and GaN[23]), shown in **Figure 4**.

This makes TiS₃ a very interesting candidate for field-effect devices requiring high current densities or high power electronics.

A deeper understanding of the breakdown process requires the analysis of the current density at breakdown ($J_{BD}$). The current density at breakdown is calculated from the current value just before breakdown and the cross-sectional area of the device, which is determined by AFM topographic characterization before breakdown. The measured breakdown current density ranges from $5 \cdot 10^4$ A cm⁻² to $1.7 \cdot 10^6$ A cm⁻² among all devices, with an average value of $6.8 \cdot 10^5$ A cm⁻². **Figure 5** shows the current density at breakdown for all 9 devices as a function of the device resistivity ($\rho$), calculated from the conductance just before breakdown. As reported in previous works,[24, 25] increasing the drain-source voltage translates into Joule heating that might be responsible for decreasing the contact resistance between the semiconducting material and the electrodes as well as annealing the material, i.e., removing adsorbates, increasing therefore the conductivity of the material, as can be depicted in the current-voltage curves shown in Figure 3a. Taking this into account, the resistivity of the material in this analysis should be calculated prior breakdown instead of at low-voltage.[24, 25]

The expected behavior of the current density as a function of the resistivity for a purely Joule heating process, in which the breakdown occurs by thermal heating, is $J_{BD} \propto 1/\sqrt{\rho}$, which arises from solving the 1D heat equation.[24] A linear fit of the experimental data ($J_{BD}$ as a function of $\rho$, red line in Figure 5) gives a slope of -0.78 with a R² value of 0.99, yielding a dependence of the current density with resistivity of the form: $J_{BD} \propto \rho^{-0.78}$. This deviation from $J_{BD} \propto 1/\sqrt{\rho}$ is related to the influence of different phenomena (defects, impurities, electron-hole pairs formation…) in the breakdown process apart from a pure Joule heating process.[24-26]

We investigate the nature of the breakdown process by estimating the temperature along the device using the analytical solution to the 1D heat equation[4, 27]:

$$T(x) = T_0 + \frac{p'_x}{g}\left(1 - \frac{\cosh(x/L_H)}{\cosh(L/2L_H)}\right) \qquad (2)$$

where $T_0 = 300$ K and $L_H = (kA/g)^{1/2}$ is the characteristic healing length along the material, $p'_x \approx I^2R/L$ is the Joule heating rate in watts per unit length and $g \approx g_{ox} \, g_{Si} \, /[L(g_{ox} + g_{Si})]$ is the thermal conductance to the substrate per unit length (with $g_{ox} = k_{ox}WL/t_{ox}$ and $g_{ox} = k_{Si}(WL)^{1/2}$, where $t_{ox}$ is the SiO₂ thickness (285 nm)). Using equation (2) for the TiS₃ devices (with values $k_{TiS3} = 3.6$ W·K⁻¹ m⁻¹,[28] $k_{ox} = 1.4$ W·K⁻¹ m⁻¹, $k_{Si} = 50$ W·K⁻¹ m⁻¹) we find that the critical temperature at the centre of the nanoribbons is ~ 400 ± 50 °C (**Figure 6**).

We study the stability and the plausible generation of defects in TiS₃ at the temperatures reached right before breakdown by performing a thermogravimetric coupled to mass spectrometry analysis of bulk TiS₃ both under Ar or O₂ rich atmosphere. The thermogravimetric analysis (TGA) of TiS₃ in an O₂ rich atmosphere shows a total mass release of 36 % from room temperature up to 650 °C (**Figure 7a**). We distinguish two events in the TGA analysis related to the mass release: one event occurs in a temperature range between 300 °C and 450 °C, exhibiting an important mass release of ~18% related to the presence of O₂, and the second event occurs in a temperature range between 450 °C and 550 ºC, with a mass release of ~18%. The total amount of mass released, as well as the stoichiometry ratio of the decomposed sample (S/Ti = 1.9 ± 0.1 and O/Ti = 2.0± 0.2) obtained by energy-dispersive X-ray spectroscopy (EDX) measurements, indicate that TiS$_3$ is decomposed into TiS$_2$ and TiO$_2$, in agreement with previously reported studies.[29] In fact, mass-spectrometry (**Figure S3** of the Supporting Information) confirms that O$_2$ is starting to be consumed during this event, supporting a reaction between TiS$_3$ and molecular O that forms an oxysulfide (TiS$_{3-x}$O$_x$) as an intermediate step on TiO$_2$ formation.[30] Higher temperatures lead to the decomposition of the remaining TiS$_3$ into TiS$_2$ as confirmed by X-ray diffraction analysis (XRD). XRD-Pattern shows the decomposition of the initial monoclinic TiS$_3$ (Figure 7b, blue line) into three crystalline phases (Figure 7b, red line): hexagonal titanium disulfide (TiS$_2$), and titanium dioxide (TiO$_2$, anatase and rutile), supporting the reaction of O with TiS$_3$. Note that similar thermal events are observed even when the TGA is performed in argon atmosphere (inset of Figure 7a) with a small amount of residual O$_2$.

These experimental results are compared with a series of DFT calculations related to the loss of S and the influence of O in the transformation from TiS$_3$ into TiS$_{3-x}$O$_x$. On the basis of our DFT calculations, we have identified several active desorption processes that may take place at the experimental temperatures (**Figure 8**): (a) desorption of S atoms, (b) desorption of S atoms due to the interaction with O atoms in the form of SO molecules that leads to the creation of a mono-vacancy, and (c) the creation of a di-vacancy. In order to quantitatively study these desorption and adsorption processes, we have used a 3x3 supercell with 72 atoms, as shown in Figure 8 (details on the calculations can be found in the Supporting Information).

We evaluate, in the first place, the viability of the oxidation of TiS$_3$ surface (before addressing the three processes mentioned above). To this aim we calculate the binding energy of an O atom adsorbed on an S atom from the equation $E_b = 1/N_O[E_{TiS_3+O} - (E_{TiS_3} + N_O E_{O_2}/2)]$, where $N_O$ is the number of O atoms in the supercell used in the calculation, $E_{TiS_3+O}$ is the total energy of O-doped TiS$_3$, $E_{TiS_3}$ is the total energy of pristine TiS$_3$ and $E_{O_2}$ is the total energy of the O$_2$ (triplet state) molecule. We obtain a binding energy of -3.74 eV, for our minimum concentration of one O atom per supercell, which is in good agreement with previous theoretical work.[31] The binding energy for other concentrations is shown in **Table 1**. According to the definition above, a negative $E_b$ indicates that the chemisorption is exothermic (energetically favored) for all possible concentration values. Except for the minimal concentration, the binding energy $E_b$ does not appreciably change with this or with the actual distribution of O (see Table 1). The actual dissociation of O$_2$ and accompanying adsorption is an activated process, but the activation energy is smaller than the experimental temperatures for all the paths that we have explored (not shown).

Since it looks that the oxidation of the surface is favorable, we now take into account the possibility of desorption of SO pairs. The formation energies of the vacancies were obtained from $E_f = E_v - E_M + E_A$, where $E_A$ is the energy of the removed isolated atom/pair (S/SO), and $E_M$ and $E_v$ are the total energies of the defect-free and the vacancy-containing layer. The different processes are represented in Figure 8, which also depicts the initial (side view) and final (top view) structures. The formation energy of an S-vacancy is 3.33 eV (Figure 8a) and the structure clearly shows locally reconstructed bonds with the nearest Ti atoms with a bond length of 2.31 Å. The formation energy is almost as high as the energy needed to desorb an O atom from S (3.67 eV), thus both processes will be excluded from our simulations since they are not active at the experimental temperatures. The energy needed to remove a SO pair from our (50 % oxidized) surface is however much lower: 1.50 eV (Figure 8b). After the relaxation, the two nearest neighbor SO-pair present a small distortion around the vacancy, which translates into a formation energy that increases to 1.63 eV when creating a second SO vacancy next to an existing one (not shown). The energy to remove the second SO pair from the same (bridge) site to create a di-vacancy (Figure 8c) is 2.72 eV. These results are in good agreement with the activation energies obtained from the non-isothermal Kissinger method[32] (**Figure S4** of the Supporting Information) using the TGA measurements in Figure 7a, where we find an activation energy of $E_{a,1} = 1.5$ eV for the first event in the TGA, i.e., SO-vacancy formation.

As far as the structural properties at higher temperature is concerned and in order to get some qualitative insight into the formation of TiS₂, we have computed the surface formation energy in the absence and presence of vacancies. The energy required to create a new surface after the split-off of one TiS₃ layer (half of the fundamental Ti₂S₆ unit) is 2.2 eV per cell. The presence of vacancies on the topmost TiS₃ layer does not hinder this process and the energy cost for a detachment of this surface layer (now TiS₃₋ₓ) is similar in energy. Specifically, for a full vacancy coverage (one S missing atom per bridge as discussed above) leading to a detached TiS₂ layer we obtain 2.35 eV. These results are in good agreement with the second activation energy, as shown in Figure S4b, where we find an activation energy of $E_{a,2} = 2.05$ eV for the second event which leads to the TiS₂ formation.

For investigating the dynamics of S desorption from the surface of single layer TiS₃, we have implemented an object kinetic Monte Carlo (OKMC) algorithm that includes a total of 4 events. However, the event that denotes the formation of the S-vacancy in the single layer of TiS₃ does not interfere in the desorption process due to its high activation energy (3.33 eV), yielding a higher probability for the desorption of the SO-pair, since it is more energetically favorable (1.50-1.63 eV). Thus, the events that actively participate in the desorption process are: SO mono-vacancy with energies 1.50 and 1.63 eV (these events occur almost simultaneously due to the low energy difference between them) and SO di-vacancy (2.72 eV). As a result, we show the evolution of the relative desorption of SO ($\Delta m$) atoms on the surface of the TiS₃ first layer as function of temperature (**Figure 9**) and for the respective time measurement: 1 minute, 5 minutes and 10 minutes. We observe that the population of SO atoms on the surface start to decrease after ≈ 200 °C and that this first process (creation of mono-vacancies) occurs until ≈ 300 °C. The formation of di-vacancies occurs in the temperature range of 500 °C to 650 °C. These results are in good agreement with the TGA measurements shown above (Figure 7a), where the first event takes place in the temperature range between 300 °C and 400 °C, and are furthermore in agreement with the estimated temperature at breakdown for the FETs, which is ~ 400 ± 50 °C. However, the second event in the TGA measurements (formation of TiS₂) couldn't be reproduced with the simplified scenario considered in the theoretical calculations. A more thorough study of the different possible reactions at high temperatures lays out of the scope of the current manuscript.

**Conclusions**

In summary, we measured the maximum current density in TiS₃ nanoribbons-based FETs by inducing electrical breakdown in the devices, finding a maximum value of $1.7 \cdot 10^6$ A cm$^{-2}$, which is higher than the maximum current density in many other semiconducting low-dimensional materials. We have studied the thermal mechanisms that may be involved in the electrical breakdown by TGA measurements in thin films, as well as by DFT calculations combined with kinetic Monte Carlo, finding that the creation of vacancies due to the oxidation of the material and the subsequent desorption of sulfur atoms could trigger the degradation and therefore the breakdown of the material in FETs. These results do not only make TiS₃ a prospective material for high power nanoelectronics requiring to stand high current densities, but also provide some insight into the thermal mechanisms occurring in nanostructured devices based on this material.

## Experimental section

*Synthesis of TiS₃ powder:* formation of sulfides is accomplished by a solid-gas reaction between titanium powder (Goodfellow, 99.5% purity) and sulfur powder by heating of elemental sulfur powder (~ 1.4 g, Merck, 99.99% purity) into vacuum sealed ampoules at 500 ºC (S/Ti molar ratio = 3) during 10 days. The as-grown ribbons present lengths of hundreds of microns, widths of a few microns and thicknesses of hundreds of nanometers, and they can be easily produced in grams scale with conventional laboratory equipment although when they are exfoliated in order to fabricate devices, the thickness can be decreased to 10-30 nm while the length and width remain almost unaltered.

*Field-effect transistors fabrication and electronic characterization:* the field-effect transistors based on single TiS₃ nanoribbons were fabricated by mechanical exfoliation from a TiS₃ thin film with polydimethylsiloxane (PDMS) stamps (Gelfilm from Gelpak) and transferred to a SiO₂/Si(p⁺) substrate with a SiO₂ thickness of 285 nm, by deterministic transfer.[15] Standard electron-beam lithography was used to fabricate the Ti/Au (5 nm/50 nm) electrodes. The atomic force microscopy (AFM) characterization of the devices was performed in a Digital Instruments D3100 AFM. The electronic characterization and the electrical breakdown measurements were carried out in a Lakeshore Cryogenics tabletop probe station in air and with homemade electronics.

*Thermogravimetric analysis and mass-spectrometry:* thermal decomposition of the sulfides has been investigated by concomitant TGA-MS measurements (under an Ar flow of 15 ml/min in TGA system (mod. Q-500) coupled to a quadrupole mass spectrometer (QMS, Pfeiffer, Switzerland). Sample without O₂ exposure was maintained at RT during 15 min to purge the system. Heating rates used were 5, 10 and 20º/min from T = 30 ºC to 650ºC.

*Density functional theory calculations:* All the total energy calculations were performed within the density functional theory (DFT) using a plane-wave basis set as implemented in the Quantum Espresso package.[33] The cutoff energy for the plane waves was chosen to be 60 Ry. The Perdew−Burke−Ernzerhofs (PBE) version of generalized gradient approximation (GGA)[34] was used for the description of the exchange correlation functional. The Brillouin zone sampling was made using the Monkhorst-Pack method.[35] The layers were separated by a 15 Å vacuum to minimize interactions between the periodic images. We have therefore optimized the atomic positions with a residual force after relaxation of $10^{-3}$ atomic units and all the configurations are ionically relaxed using the Broyden–Fletcher–GoldfarbShann's procedure. From the energetics extracted by DFT calculations, we employed Kinetic Monte Carlo (KMC) calculations to understand the time evolution of the surface distribution of the vacancies for different temperatures and different vacancy-type defects on monolayer TiS₃.

## Acknowledgements

A.C-G. acknowledges financial support from the BBVA Foundation through the fellowship "I Convocatoria de Ayudas Fundacion BBVA a Investigadores, Innovadores y Creadores Culturales" ("Semiconductores ultradelgados: hacia la optpelectronica flexible"), from the MINECO (Ramón y Cajal 2014 program, RYC-2014-01406), from the MICINN (MAT2014-58399-JIN) and the European Commission under the Graphene Flagship, contract CNECTICT-604391. A.J.M-M., G.R-B. and N.A. acknowledge the support of the MICCINN/MINECO (Spain) through the programs MAT2014-57915-R, BES-2012-057346 and FIS2011-23488 and Comunidad de Madrid (Spain) through the program S2013/MIT-3007 (MAD2D). J.M.C., J.RA, E.F, F.L., C.S., I.J.F acknowledge finantial support from MINECO-FEDER (MAT2015-65203-R) and Mexican National Council for Science and Technology (CONACIT) México. J.I. and H.S.J.vd.Z acknowledge the support of the Dutch organization for Fundamental Research on Matter (FOM) and by the Ministry of Education, Culture, and Science (OCW). W.S.P acknowledges CAPES Foundation, Ministry of Education of Brazil, under grant BEX 9476/13-0. W.S.P and J.J.P acknowledge Ministerio de Economía y Competitividad (Spain) for financial support under grant FIS2013-47328-C02-1, the European Union structural funds and the Comunidad de Madrid MAD2D-CM program under grant nos. S2013/MIT-3007, the Generalitat Valenciana under grant no. PROMETEO/ 2012/011. W.S.P and J.J.P also acknowledge the computer resources and assistance provided by the Centro de Computación Científica of the Universidad Autónoma de Madrid and the RES.

Aday J. Molina-Mendoza and Joshua O. Island contributed equally to this work.

## Figures

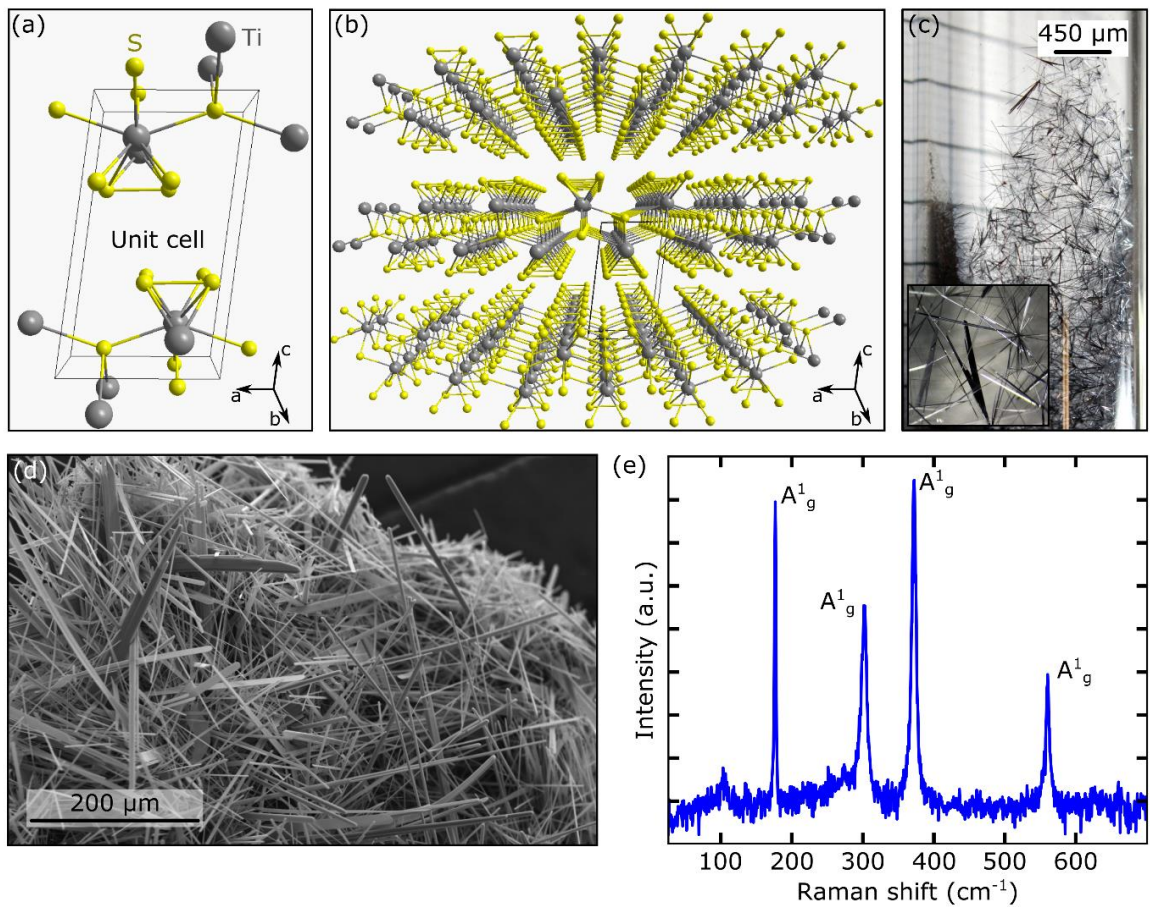

**Figure 1. (a)** Artistic representation of the TiS₃ unit cell where the grey spheres represent the Ti atoms and the yellow spheres represent the S atoms. **(b)** Artistic representation of the layered crystal structure of TiS₃. The unit cell is indicated by solid black lines. **(c)** Photograph of TiS₃ inside the ampoule used to grow the material. Inset: magnified photograph of TiS₃ ribbons. **(d)** SEM image of the TiS₃ powder showing the nanoribbon morphology of the material. **(e)** Raman spectra acquired in the TiS₃ powder.

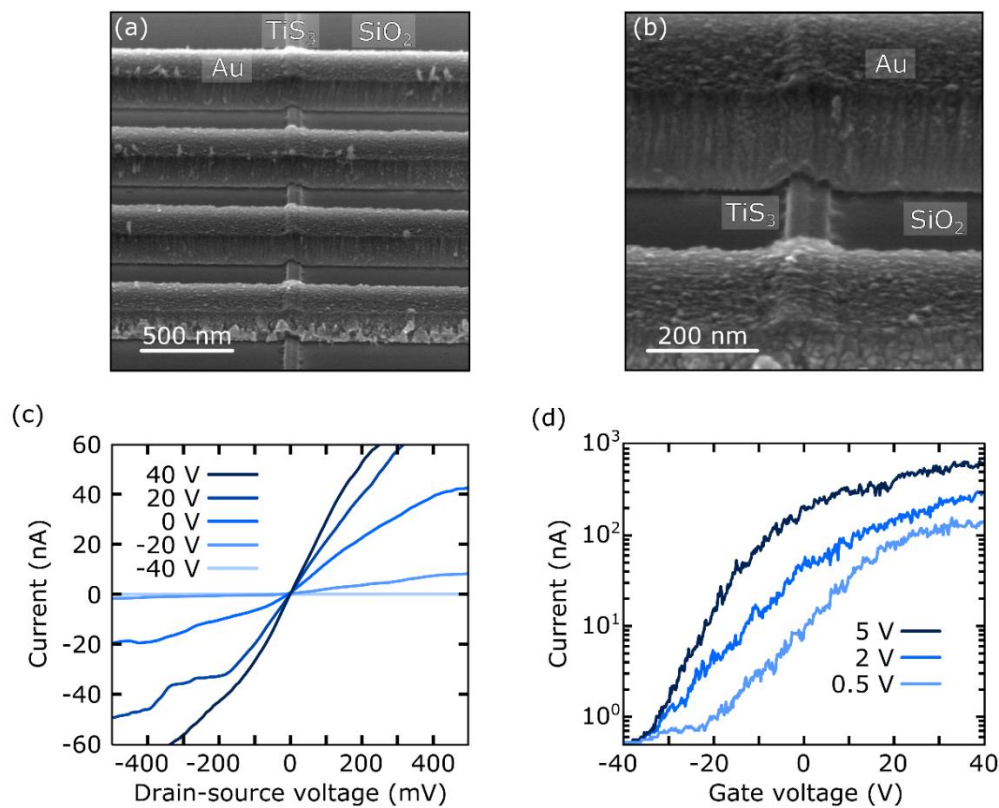

**Figure 2.** **(a)** SEM image of TiS₃ FET fabricated with from one TiS₃ nanoribbon. **(b)** Higher resolution SEM image of one of the TiS₃ FET shown in (a). **(c)** Current-voltage characteristics of a TiS₃ nanoribbon device for different back-gate voltages. **(d)** Transfer curves of a TiS₃ nanoribbon FET for different drain-source voltages.

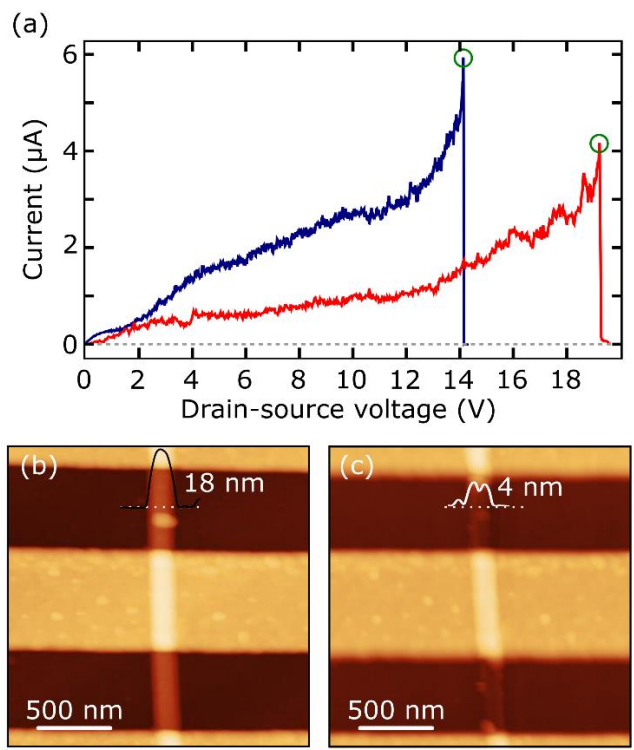

**Figure 3.** **(a)** Current-voltage characteristics during the electrical breakdown process of two TiS₃ devices. The green circles highlight the current and voltage values just before breakdown. **(b)** and **(c)** AFM topographic images of one device before (b) and after (c) electrical breakdown including the AFM line profile that indicates the thickness.

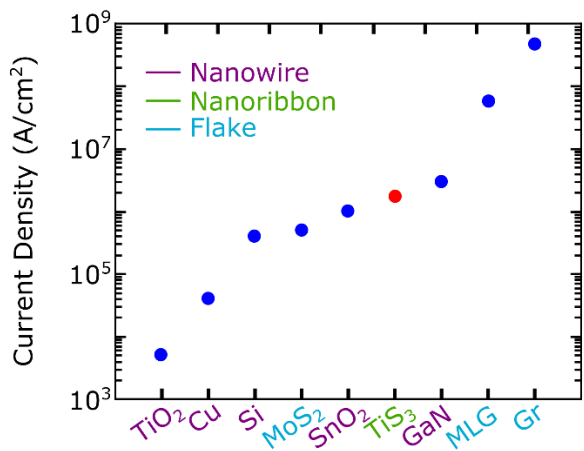

**Figure 4.** Maximum current density at breakdown for different nanomaterials: multilayer graphene,[18] monolayer graphene,[19] Cu,[20] TiO₂,[21] SnO₂,[21] Si,[22] and GaN[23] nanowires, multilayer MoS₂,[17] and TiS₃ (present work).

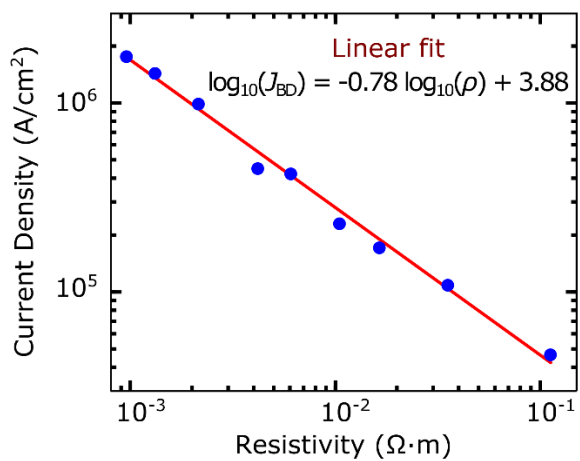

**Figure 5.** Current density at breakdown voltage versus resistivity in logarithmic scale for all the measured devices (blue dots). The red line represents a linear fit which yields a slope of -0.78.

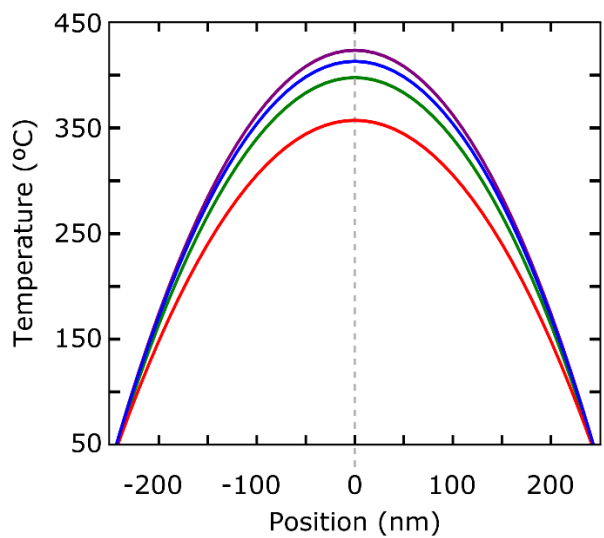

**Figure 6.** Calculated temperature distribution at breakdown along four TiS₃ nanoribbons-based devices using the 1D heat equation. The zero position (dashed grey line) corresponds to the center of the nanoribbon, while the -250 nm and 250 nm correspond to both source and drain electrodes edges. The highest temperature is reached in the center of the device and oscillates between 350 °C and 450 °C.

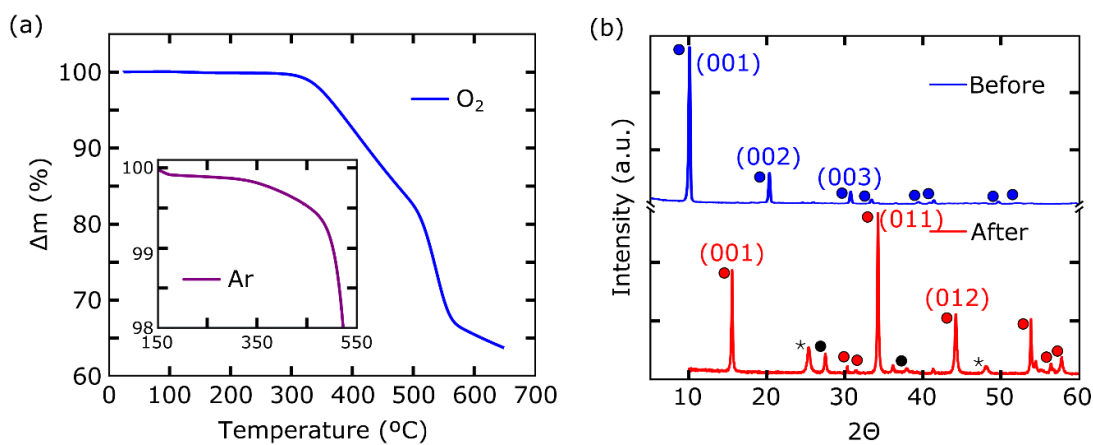

**Figure 7.** **(a)** Thermogravimetric analysis of bulk TiS₃ in oxygen atmosphere. Inset: Thermogravimetric analysis of bulk TiS₃ in argon around the first event. **(b)** XRD pattern of TiS₃ as-synthetized (blue line) and after thermal treatment (red line). The blue circles correspond to TiS₃ phase, the red circles correspond to TiS₂ phase, * correspond to TiO₂ anatase and the black circles correspond to TiO₂ rutile.

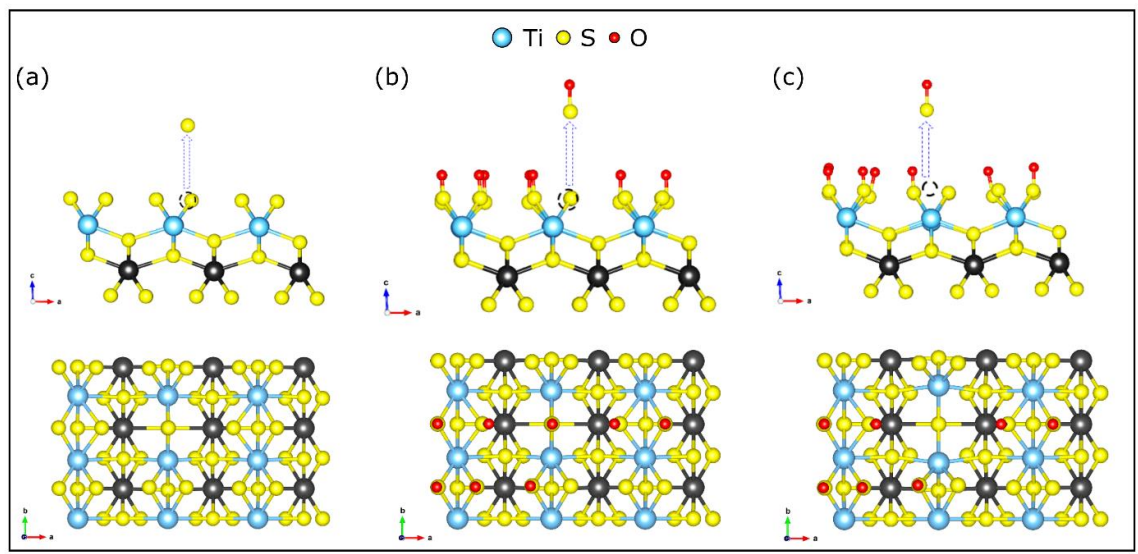

**Figure 8.** Artistic drawing of the different process considered in the DFT calculations. **(a)** S atoms desorption from the surface. **(b)** SO pairs desorption from the surface, creating a mono-vacancy. **(c)** SO pairs desorption from the surface, creating a di-vacancy.

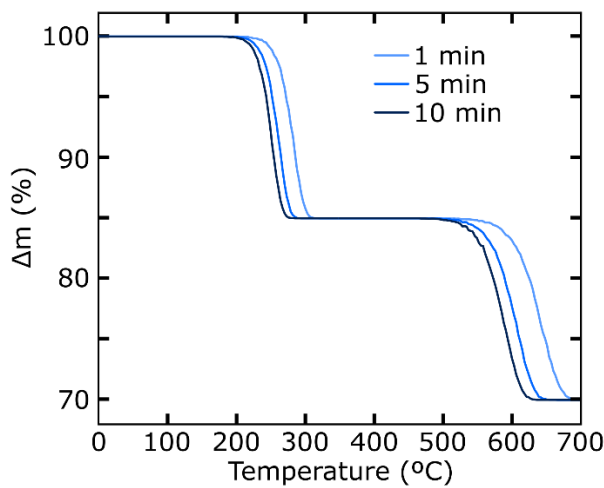

**Figure 9.** Calculated evolution of the relative desorption of SO ($\Delta m$) atoms on the surface layer of TiS₃ as a function of the temperature for different time rates: 1 minute (light blue), 5 minutes (medium-dark blue) and 10 minutes (dark blue).

**Table 1.** Binding energy for the oxygen impurities for different concentrations.

| O surface concentration (per supercell) (%) | Binding energy (eV) |
|---|---|
| 5.5 | -3.74 |
| 22.5 | -2.78 |
| 50.0 | -2.58 |
| 100 | -2.64 |

## Supporting Information

## High current density electrical breakdown of TiS₃ nanoribbon-based field-effect transistors

*Aday J. Molina-Mendoza,* [†]* *Joshua O. Island,* [†] *Wendel S. Paz, Jose Manuel Clamagirand, Jose Ramón Ares, Eduardo Flores, Fabrice Leardini, Carlos Sánchez, Nicolás Agraït, Gabino Rubio-Bollinger, Herre S. J. van der Zant, Isabel J. Ferrer, J.J. Palacios, and Andres Castellanos-Gomez.**

[†]*These authors contributed equally*

Aday J. Molina-Mendoza[1], Wendel S. Paz[1], Prof. Nicolás Agraït[1,2,3], Prof. Gabino Rubio-Bollinger[1,3] and Prof. J.J. Palacios[1]

[1]Departamento de Física de la Materia Condensada, Universidad Autónoma de Madrid, Campus de Cantoblanco, E-28049, Madrid, Spain.

[2]Instituto Madrileño de Estudios Avanzados en Nanociencia (IMDEA-nanociencia), Campus de Cantoblanco, E-18049 Madrid, Spain.

[3]Condensed Matter Physics Center (IFIMAC), Universidad Autónoma de Madrid, E-28049 Madrid, Spain

E-mail: aday.molina@uam.es

Dr. Joshua O. Island[4[+]] and Prof. Herre S. J. van der Zant[4]

[4]Kavli Institute of Nanoscience, Delft University of Technology, Lorentzweg 1, 2628 CJ Delft, The Netherlands

[+]Current Address: Department of Physics, University of California, Santa Barbara, California 93106-6105 USA

Dr. Jose Manuel Clamagirand[5], Prof. Jose Ramón Ares[5], Eduardo Flores[5], Prof. Fabrice Leardini[5], Prof. Isabel J. Ferrer[5,6] and Prof. Carlos Sánchez[5,6]

[5]Materials of Interest in Renewable Energies Group (MIRE Group), Dpto. de Física de Materiales, Universidad Autónoma de Madrid, Campus de Cantoblanco, E-28049 Madrid, Spain

[6]Instituto de Ciencia de Materiales "Nicolás Cabrera", Campus de Cantoblanco, E-18049 Madrid, Spain.

Dr. Andres Castellanos-Gomez[2]

[2]Instituto Madrileño de Estudios Avanzados en Nanociencia (IMDEA-nanociencia), Campus de Cantoblanco, E-18049 Madrid, Spain.

E-mail: andres.castellanos@imdea.org

**Contact resistance from transfer length measurement**

Figure S1 shows an AFM scan of the device used to estimate the contact resistance using the transfer length method (TLM). All possible combinations of two electrodes were used in order to extract 10 data points of resistance as a function of channel length. From a linear fit of the data in Figure S1b, we estimate a contact resistance of 8.40 ± 4.60 MΩ/µm. The measurements were taken at $V_g = 0$ V and $V_b = 500$ mV.

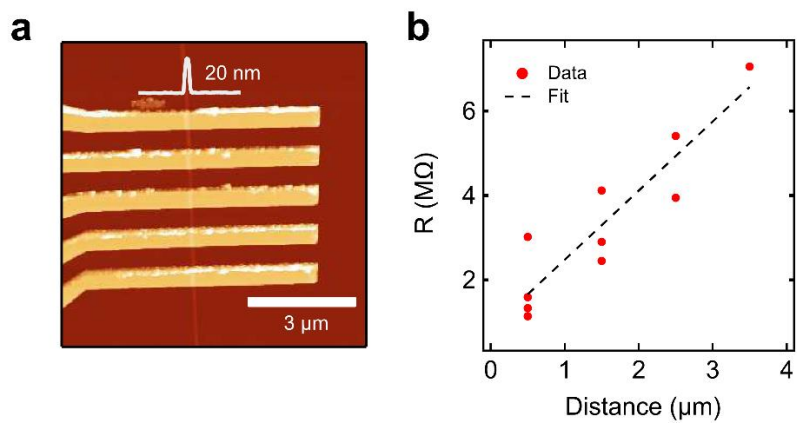

**Figure S1.** **(a)** AFM image of the TiS₃ nanoribbon device used to estimate the contact resistance. **(b)** Plot of the two terminal channel resistance ($V_b = 500$ mV, $V_g = 0$V) as a function of channel length.

**Electrical breakdown in vacuum**

In Figure S2 we show the current-voltage trace measured during the breakdown process for one TiS₃ device in vacuum ($P < 10^{-5}$ mbar). As it can be depicted, the voltage needed to electrically break the nanodevice is in the same order of magnitude that those needed to break the nanodevices in air. The current density at breakdown measured for this device is $9.5 \cdot 10^5$ A/cm², in the same order of magnitude as the devices measured in air. This result suggests either that the air is not important in the breakdown process or that even a small quantity of oxygen that might be adsorbed on the nanoribbon surface is enough to trigger the electrical breakdown. As it will be discussed later from the TGA measurements it is found that the oxygen adsorbed on the surface is playing an important role in the process.

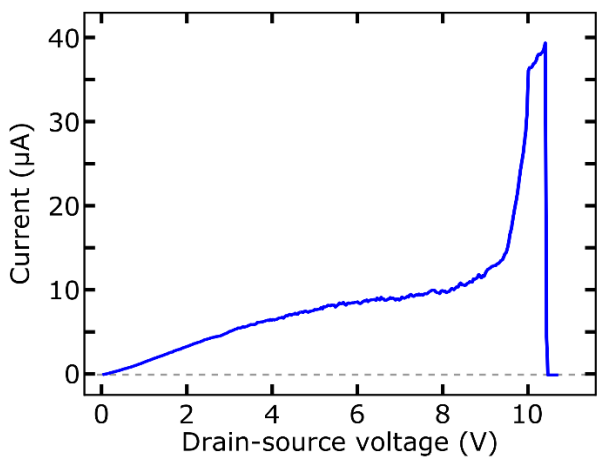

**Figure S2.** Current-voltage curve of a TiS₃ nanoribbon-based device measured in vacuum ($P < 10^{-5}$ mbar). The electrical breakdown occurs at a current density of $9.5 \cdot 10^5$ A/cm², in the same range as the devices measured in air.

## Mass-spectrometry during TiS₃ thermal decomposition

The mass-spectrometry analysis of TiS₃ during the thermal treatment at m/q = 32, attributed to O₂ consumption, is shown in Figure S3. It is observed that the signal (current) decreases at a small rate while increasing the temperature from room temperature to 300 °C, but it decreases with a higher rate between 300 °C and 550 °C, indicating a higher oxygen consumption in this temperature range, related to a reaction between O₂ and TiS₃.

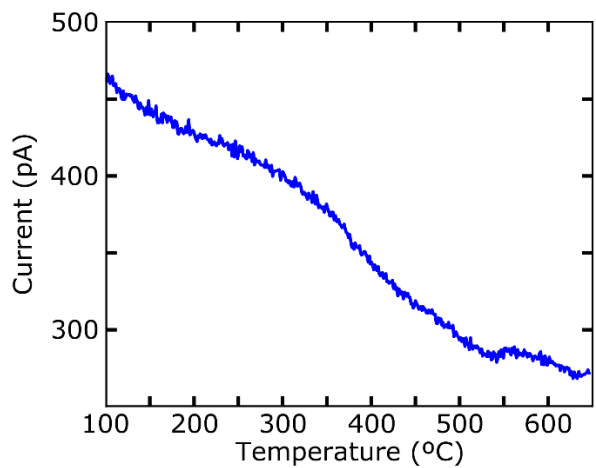

**Figure S3.** Ion current signal related to molecular oxygen (m/q = 32) during thermal decomposition of TiS₃. An increase of oxygen consumption from T = 350 °C up to 550 °C is observed (O₂- ion current signal decreases) which is related to the reaction between oxygen and TiS₃.

### Density functional theory calculations of activation energies

We study from the density functional theory (DFT) and kinetic Monte Carlo (KMC) simulations the thermal stability of the single layer TiS₃ on the atmosphere with low concentration of the atomic O. We have calculated the activation energies for the desorption of S atoms, SO pair and O atoms on the surface of a single layer of the TiS₃ and TiS₃O$_x$.

We have calculated the total energies of a single layer TiS₃, TiS₃ with a vacancy and TiS₃O$_x$ with SO pair vacancy using the Quantum Espresso package. [33] By starting from the initial lattice parameters extracted in this paper we have built a 3 × 3 supercell to reduce the interaction between vacancies on surface TiS₃. The layers were separated by a 15 Å (vacuum) to minimize interactions between the periodic images. We have therefore optimized the atomic positions with a residual force after relaxation of 0.001 atomic units, for the pristine system, TiS₃ with a vacancy, TiS₃O$_x$ and TiS₃₋$_x$O$_x$. All the configurations are ionically relaxed using the Broyden–Fletcher–GoldfarbShann's (BFS) procedure. The generalized gradient approximation of Perdew-Burke-Ernzerhof (GGA-PBE)[34] was adopted for exchange-correlation functional, and the Trouiller-Martins norm-conserving pseudo-potential for Ti, S and O is used to model valence electron–nuclei interactions. The energy cut-off for the plane wave basis set is put at 60 Ry with a charge density cut-off of 240 Ry.

The KMC calculations have been performed to understand the time evolution of the surface distribution of the vacancies for different temperatures and different vacancy-type defects on monolayer TiS₃. KMC algorithms are powerful techniques to study the dynamics of a system of particles when the different events that those particles can perform are known as well as their probabilities.[36] We use an object kinetic Monte Carlo (OKMC) algorithm, based on the residence time algorithm or Bortz-Kalos-Liebowitz (BKL) algorithm.[36, 37] Typically in OKMC simulations, a list of possible events is defined with a given probability for each event, $\Gamma_i$. This probability usually follows an Arrhenius dependence with temperature:

$$\Gamma_i = \Gamma_0 \exp\left(\frac{-\Delta_i}{K_B T}\right) \text{(1)}$$

where $K_B$ is the Boltzmann constant, $\Delta_i$ is the activation energy of the given event, and $\Gamma_0$ is the attempt frequency (we assume a reasonable attempt frequency $\Gamma_0 = 10^{13}/s$). Here, the activation energies are related to substitutional and desorption energies.

The total rate for all events, R, is then calculated as

$$R = \sum_{i=1}^{n_e} \Gamma_i \, N_i \text{(2)}$$

where $n_e$ is the total number of events and $N_i$ is the number of particles that can perform event i. An event is then selected by picking a random number between [0, R]. In this way, one event is selected every Monte Carlo step from all possible with the appropriate weight. Once the event has been selected, a random particle is chosen from all those that can undergo that event. The particle is then moved and the total rate has to be computed again for the next simulation step. At every Monte Carlo step, the time increases by

$$t = -ln\frac{\xi}{R} \text{(3)}$$

where $\xi$ is a random number between [0, 1] that is used to give a Poisson distribution of the time.

**Experimental activation energies**

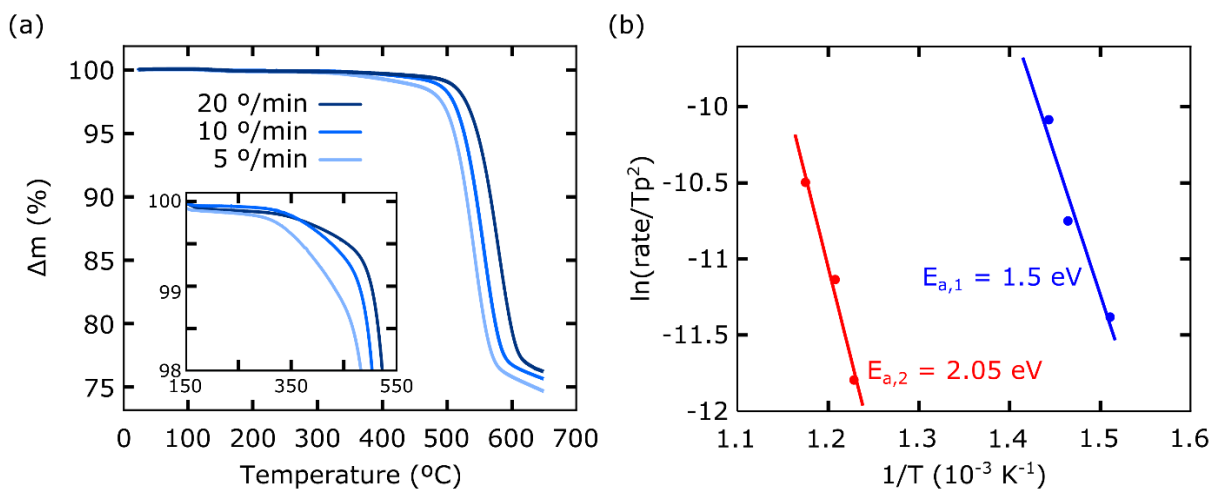

**Figure S4.** **(a)** Thermogravimetric curves obtained at different heating rates of TiS₃ under argon atmosphere. **(b)** Kissinger plots and calculated activation energies of the two events.